# Estimation of turbulent heating of solar wind protons at 1AU

G. Livadiotis[1], M. A. Dayeh[1,2], G. P. Zank[3,4]

1 Space Science and Engineering Division, Southwest Research Institute, San Antonio, TX, USA
2 Department of Physics and Astronomy, University of Texas at San Antonio, San Antonio, TX, USA
3 Center for Space Plasma and Aeronomic Research (CSPAR), University of Alabama in Huntsville, Huntsville, AL 35899, USA
4 Department of Space Science, University of Alabama in Huntsville, Huntsville, AL 35899, USA

## Abstract

The paper presents a new method for deriving turbulent heating of the solar wind using plasma moments and magnetic field data. We develop the method and then apply it to compute the turbulent heating of the solar wind proton plasma at 1AU. The method employs two physical properties of the expanding solar wind plasma, the wave-particle thermodynamic equilibrium, and the transport of entropic rate. We analyze plasma moments and field datasets taken from Wind S/C, in order to compute (i) the fluctuating magnetic energy, (ii) the corresponding correlation length, and (iii) the turbulent heating rate. We identify their relationships with the solar wind speed, as well as the variation of these relationships relative to solar wind and interplanetary coronal mass ejection plasma.

## 1. Introduction

We employ two physical properties of the expanding solar wind plasma for improving our understanding of the turbulent heating of solar wind at 1AU, i.e., (a) wave-particle thermodynamic equilibria (Livadiotis 2019a), and (b) transport of the rate of entropy change (Adhikari et al. 2020).

Recently, it was shown that the energy transfer between plasma particles and waves is governed by a new and unique relationship: the ratio between the energy per particle over the plasma frequency is constant, that is, a large-scale analogue of Planck's constant denoted by $\hbar_*$ (Livadiotis & McComas 2013, 2014a, 2014b; Witze 2013; Livadiotis 2015, 2016, 2017, Ch.5; 2019a; Livadiotis & Desai 2016; Livadiotis et al. 2018). As an example, Figure 1 demonstrates the large variation of the representative average values and uncertainties of the plasma parameters of 27 space and astrophysical plasmas (Livadiotis 2019a), which are stable and residing at stationary states (Livadiotis 2018a;b), while the respective values of the ratio $E_p/\omega_{pl}$ that approaches the value of $\hbar_*$ remains almost constant.

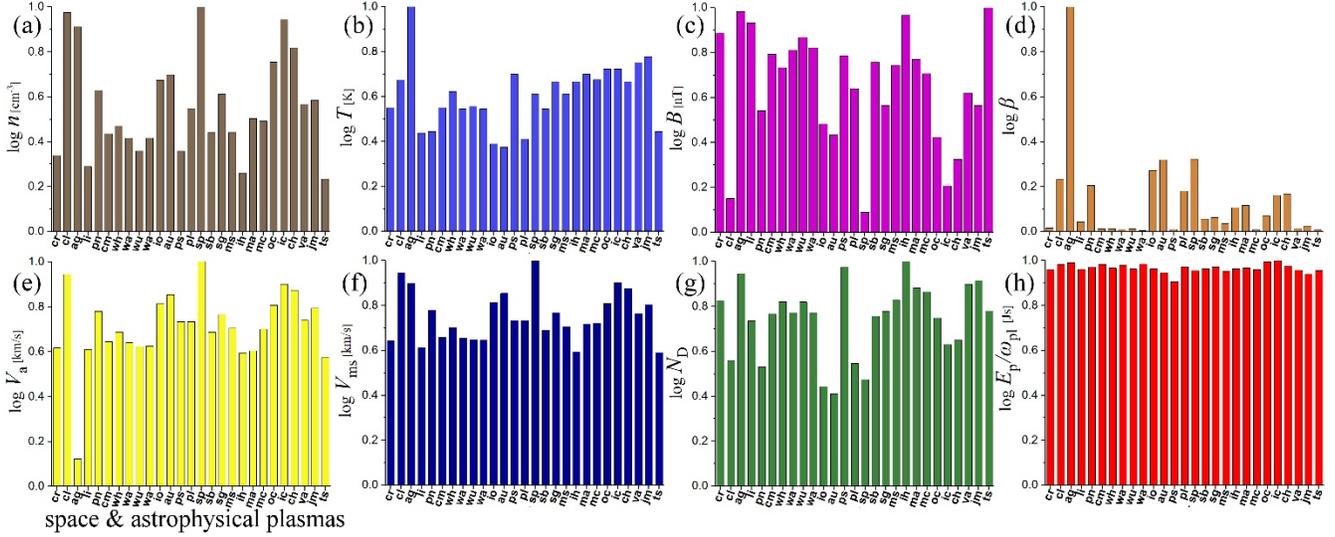

**Figure 1.** Average logarithm values of various plasma parameters and of the ratio $E_p/\omega_{pl}$ for 27 different space and astrophysical plasmas (stable and residing at stationary states), where the variation of all the plotted parameters in contrast to the constancy of $E_p/\omega_{pl}$ is apparent. The plotted color-coded parameters are: (a) density (grey), (b) temperature (light blue), (c) magnetic field strength (magenta), (d) plasma beta (brown), (e) Alfvén speed (yellow), (f) fast magnetosonic speed (deep blue), (g) Debye number (number of particles in a Debye sphere) (green), and (h) the ratio of the energy per proton over the plasma frequency, $E_p/\omega_{pl}$ (red). All parameter values are normalized to 1 (by dividing each of the 27 values with the maximum between them). The 27 plasmas and their abbreviation is the following (from left to right at each panel): CIRs (cr), Coronal Loops (cl), AGN (ag), LISM (li), Planetary Nebula (pn), CMEs (cm), Solar Wind – Helios (wh), Solar Wind – ACE (wa), Solar Wind – Ulysses (wu), Solar Wind – 1 au average (wa), Ionosphere (io), Aurora (au), Plasma Sheet (ps), Plasmasphere (pl), Sunspot Plume (sp), Shock example (by Burlaga & King 1979) (sb), Shock example (by Gopalswamy & Yashiro 2011) (sg), Magnetosheath (ms), Inner Heliosheath (ih), Magnetosphere – average (ma), Magnetosphere – Cluster (mc), Outer Corona (oc), Inner Corona (ic), Coronal Holes (ch), Van Allen Belts (va), Jovian Magnetosphere – average (jm), Termination Shock (ts); (Taken from Livadiotis 2019a)

The wave-particle thermodynamic equilibrium requires that the energy $E_{pl}$ of a plasmon (that is, the quantum of plasma oscillation) be balanced by the energy per proton $E_p$, i.e., $E_{pl}=E_p$, with:

- *plasmon energy $E_{pl}$ = energy of quanta $\hbar_* \cdot \omega_{pl}$*

- *energy per proton, $E_p$ = sum of magnetic energy $B^2/(2\mu_0 n)$, enthalpy $\gamma/(\gamma-1) \cdot k_B T$, and turbulent energy $E_t$.*

The plasma-field coupling in the wave-particle equilibrium constitutes a fine probe for estimating the turbulent energy $E_t$ (which is associated with the turbulent heating and other relevant parameters):

$$E_{pl} = \hbar_* \cdot \omega_{pl}, \quad E_p = E_p^0 + E_t \quad \text{or} \quad E_t = \hbar_* \cdot \omega_{pl} - E_p^0, \text{ with, } E_p^0 = \tfrac{1}{2\mu_0} B^2/n + \tfrac{\gamma}{\gamma-1} k_B T, \tag{1}$$

(fundamental plasma oscillation frequency: $\omega_{pl} = [n \cdot e^2 \varepsilon_0^{-1}(m_e^{-1}+m_p^{-1})]^{1/2}$; $m_e$, $m_p$: electron & proton masses; $e$: electron charge; $\varepsilon_0$: permittivity; $\mu_0$: permeability; $k_B$: Boltzmann constant; $\gamma=5/3$).

Here we use the method provided by the wave-particle equilibrium shown in Eq.(1) to derive the turbulent energy $E_t$ at $R=1$AU with respect to solar wind speed. On the other hand, the turbulent energy is connected with the turbulent heating rate through the formalism of the transport of entropy and its rate of change.

The turbulent heating $S_t$ is necessary for the entropy derivation (Adhikari et al. 2020); the equation that connects the entropy $S = \tfrac{3}{2}k_B \cdot \ln(P/n^\gamma) + const.$ with the turbulent heating $S_t$ is given by

$$dS/dR = k_B \cdot S_t / (PV_{sw}), \qquad (2)$$

where $P = nk_B T$ is the solar wind proton plasma thermal pressure; $n$ and $T$ are the proton density and temperature; $V_{sw}$ is the solar wind speed.

The turbulent magnetic energy and correlation length affect $E_t$ and $S_t$ according to (Adhikari et al. 2020),

$$E_t = m_P \cdot [E_b^0 + E_b^{A0} \log(f/f^0)] \quad \text{with} \quad f \equiv E_b^A \cdot n^{-\tfrac{1}{2}}, \; f \equiv f(R), \; f^0 \equiv f(R_0), \qquad (3a)$$

$$\tfrac{1}{n} S_t = m_P \cdot [E_b^{0\tfrac{3}{2}}/\lambda_b^0 + \sqrt{2}(E_b^{A0\tfrac{3}{2}}/\lambda_b^{A0}) \cdot (g/g^0)] \quad \text{with} \quad g \equiv (\lambda_b^A)^{-1}(E_b^A)^{-\tfrac{1}{2}}, \; g = g(R), \; g^0 = g(R_0), \qquad (3b)$$

where the different quantities are: (a) the quasi-2D $E_b^0$, and "Slab" $E_b^A$, fluctuating magnetic energies (in unit mass), and (b) their corresponding correlation lengths, $\lambda_b^0$ and $\lambda_b^A$, respectively.

Figure 2 plots the observed and/or modeled values of (a) fluctuating magnetic energies, (b) correlation lengths, and (c) entropy, as functions of the heliocentric distance $R=1$–$75$AU. The observed values were derived from Voyager 2 measurements (Adhikari et al. 2017; 2020); the modelled values were derived from a system of coupled solar wind and turbulence transport model equations (Livadiotis 2019a; Adhikari et al. 2017; Zank et al. 1996, 2012).

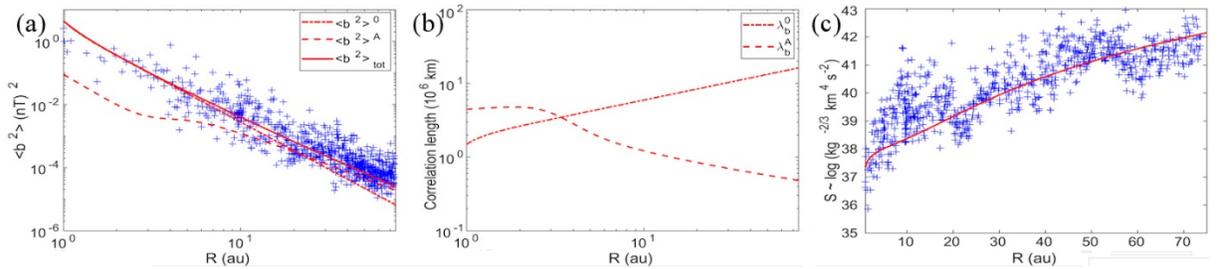

**Figure 2.** Observed (blue) and theoretical (red) values of (a) fluctuating magnetic energies $<b^2>^{0\{A\}} = E_b^{0\{A\}}\mu_0\rho$, (b) correlation lengths, and (c) entropy $S$ (with an arbitrary choice of the constant, so that $S>0$), plotted as a function of the radial distance $R$. (Taken from Adhikari et al. 2020)

In this paper, we reverse Eq.(2), in order to determine the turbulent heating $S_t$ from the entropic rate $dS/dt$, where $V_a \cdot d/dR = d/dt$ (advection speed $V_a$ = flow speed $V_{sw}$), namely,

$$S_t = P \cdot \tfrac{d}{dt}(\tfrac{1}{k_B} S) = \tfrac{3}{2} P \cdot \tfrac{d}{dt} \ln(P/n^\gamma), \text{ or } \tfrac{1}{n} S_t = \tfrac{3}{2} k_B T \cdot \tfrac{d}{dt} \ln(P/n^\gamma) = \tfrac{3}{2} k_B T \cdot \tfrac{d}{dt} \ln(k_B T/n^{\gamma-1}). \qquad (4)$$

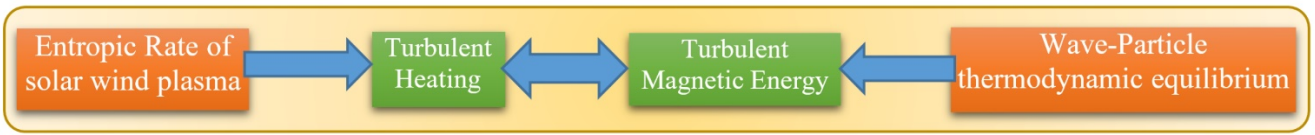

**Figure 3.** Methods (orange) for computing turbulent heating and relevant turbulence parameters (green).

In this paper, we determine the turbulent heating rate $S_t/n$, and relevant turbulence parameters, such as, the fluctuating magnetic energy $E_{b0}$, and its correlation length $\lambda_{b0}$, by combining the two methods (as shown in Figure 3): (i) wave-particle thermodynamic equilibrium, characterized by the equation connecting the proton and plasmon energies, and (ii) entropy evolution, characterized by the equation connecting the entropic rate with turbulent heating. Section 2 describes the data used in this analysis. Section 3 develops the method for deriving the turbulent heating and its related parameters; in particular we derive the relationships of turbulent heating rate with entropic rate, wave-particle equilibrium energy, and with solar wind speed; we also study the variations of turbulent heating rate and related parameters between the solar wind plasma and an interplanetary coronal mass ejection (ICME). Finally, Section 4 summarizes the conclusions, highlighting the analysis outcome, that is, a method for computing the turbulent heating using simply the plasma moments and magnetic field, and for improving our understanding of the turbulence complexity. (For instance, note that several attempts were made for the description of the complexity of turbulence, based on a new time domain, e.g., Varotsos et al. 2006, 2014, the theory of kappa distributions, e.g., Gravanis et al. 2019, and their connection with the phenomenon of wave-particle thermodynamic equilibria). The appendix demonstrates a toolbox for the derivation of the uncertainty involved in the analysis.

## 2. Data

We use ~92-second resolution of long-term observations of solar wind plasma moments (speed $V_{sw}$, density $n$, and temperature $T$ or thermal speed $V_{th} = (2k_B T / m_p)^{1/2}$) in conjunction with simultaneous measurements of the IP magnetic field strength $B$, taken from Solar Wind Experiment (SWE) (Ogilvie et al. 1995) and Magnetic Field Investigation (MFI) (Lepping et al. 1995) onboard Wind S/C, publicly accessible at http://science.nasa.gov/missions. We focus on the first 70 days of 1995 (~66000 data samples; see Figure 4) in order to show our method for deriving the turbulent heating. This time period occurred during the declining phase of solar activity cycle 23, and was characterized by ICMEs, which is apparent in increases of the solar wind density and magnetic field magnitude that precede the arrival of the high speed streams at 1 AU. ICMEs are identified using in-situ measurements of magnetic field, solar wind ions, suprathermal electrons, energetic protons, heavy ion composition, cosmic rays (Gosling 1996, 1997; Neugebauer & Goldstein 1997; Dayeh et al. 2017, 2018). ICMEs with high speeds may drive an IP shock, identified as sharp increases in speed, density, temperature, & magnetic field strength (Szabo 1994). (ICME lists are regularly compiled based on Wind & ACE observations; for ICME lists, see: Cane & Richardson 2003; Richardson & Cane 2010; see also: Jian et al. 2006a, 2006b, 2009; Kilpua et al. 2015).

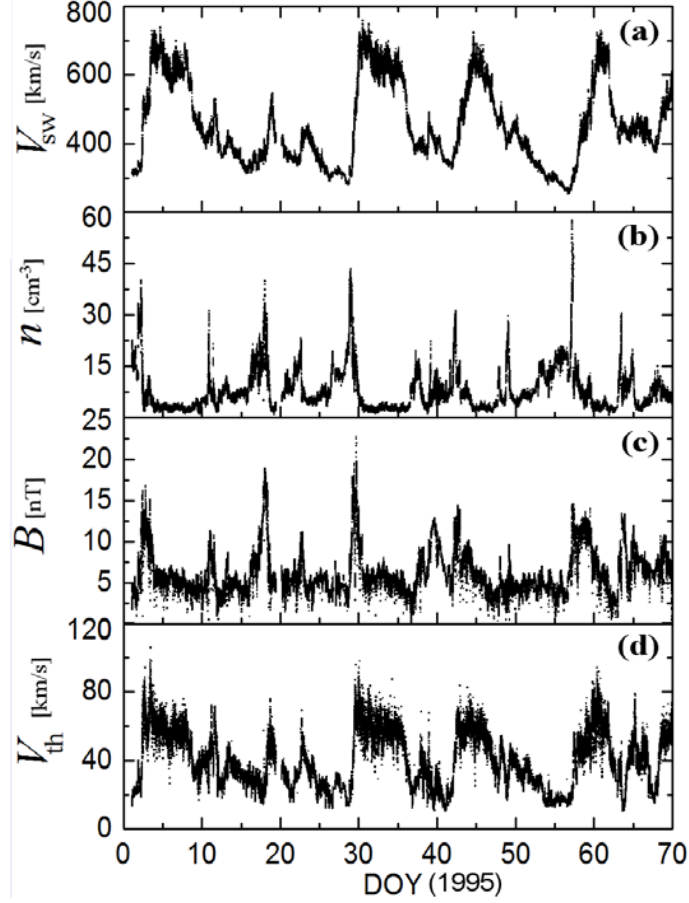

**Figure 4.** ~92-s resolution measurements of the bulk solar wind plasma parameters and magnetic field magnitude taken from SWE and MAG onboard Wind S/C during the first 70 day period of 1995.

### 3. Results: Derivation of turbulent heating parameters

#### 3.1. Relationship between turbulent heating and entropic rate

This analysis focuses on $R\sim1$ AU, hence, Eq.(3b) becomes $\frac{1}{n}S_t \cong m_p \cdot (E_b^{0\frac{3}{2}}/\lambda_b^0 + \sqrt{2}E_b^{A0\frac{3}{2}}/\lambda_b^{A0})$. The 2$^{nd}$ term in Eq.(3b) is much smaller than the 1$^{st}$ term and can be ignored in the inner heliosphere and especially for $R\sim1$ AU; indeed, as shown in Figure 2a,b, $E_b^0 \gg E_b^{A0}$ and $\lambda_b^0 \ll \lambda_b^{A0}$ so that $E_b^{0\frac{3}{2}}/\lambda_b^0 \gg E_b^{A0\frac{3}{2}}/\lambda_b^{A0}$; then, for $R\sim1$ AU, we obtain $\frac{1}{n}S_t \cong m_p \cdot E_b^{0\frac{3}{2}}/\lambda_b^0$; ($E_b^0$ is given in unit mass).

First, we use plasma moments $n$, $T$, and $P=2nk_BT$, to derive the entropy $S$, then, the entropic rate $dS/dt$, followed by the turbulent heating term $S_t$; then, we calculate $E_b^{0\frac{3}{2}}/\lambda_b^0$, i.e., the fraction involving the fluctuating magnetic energy $E_b^0$ and its correlation length $\lambda_b^0$. Therefore, we consecutively find:

$$\text{(i) } \tfrac{1}{k_B}S = \tfrac{3}{2}\ln(P/n^\gamma),\ \text{(ii) } \tfrac{1}{n}S_t = k_B T \cdot \tfrac{d}{dt}(\tfrac{1}{k_B}S),\ \text{(iii) } E_b^{0\frac{3}{2}}/\lambda_b^0 \cong m_p^{-1} \cdot \tfrac{1}{n}S_t \ . \quad (5)$$

Figure 5 shows the values of entropy $S$, entropic rate $dS/dt$, and turbulent heating term $S_t$ and rate $S_t/n$, corresponding to the data shown in Figure 4. We examine the relationships of these quantities with the solar wind speed $V_{sw}$. First, we examine the entropic rate plotted against $V_{sw}$.

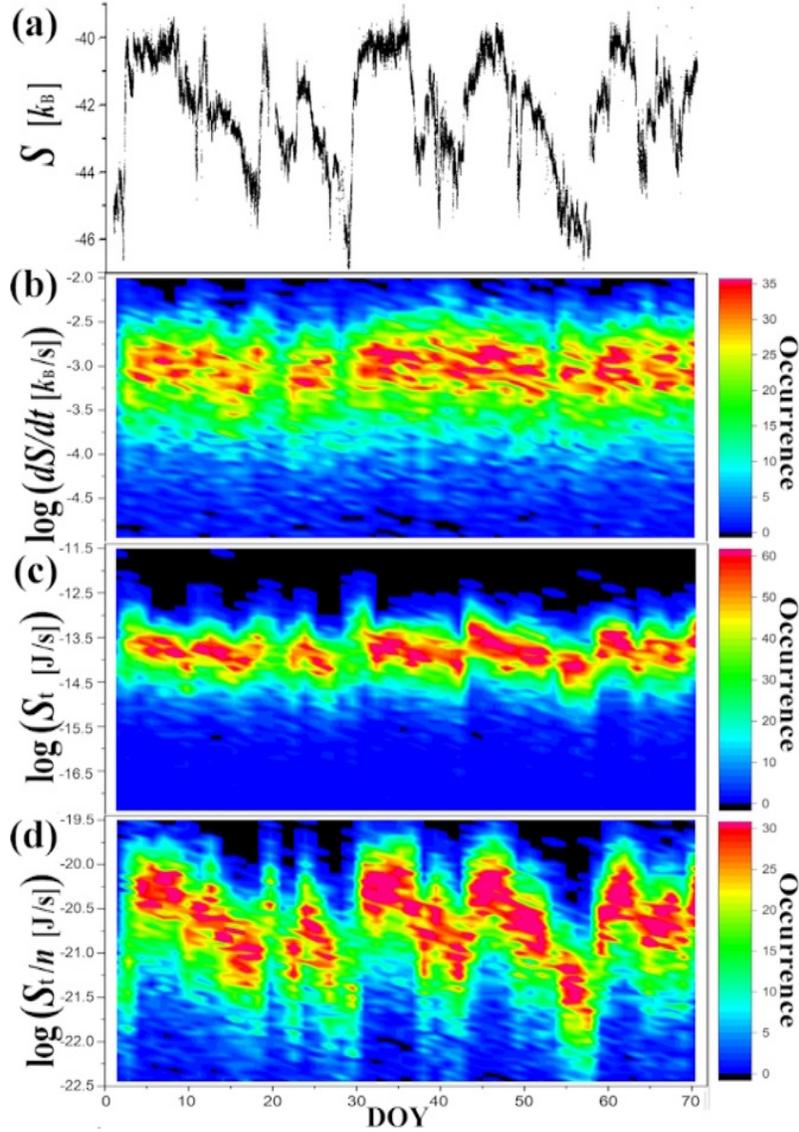

**Figure 5.** (a) Entropy $S$, (b) entropic rate $dS/dt$, (c) turbulent heating term $S_t$, (d) turbulent heating rate $S_t/n$, determined using the datasets in Figure 4; (panels (b-d) are on semi-log-scales).

Figure 6(a) displays the occurrence frequency of the solar wind speed $V_{sw}$, observed during the 70-day period shown in Figure 4(a), while Figure 6(b) shows the 2D-occurrence frequency of the entropic rate (plotted by its logarithm) taken from Figure 5(b), and plotted against $V_{sw}$. Figure 6(a) demonstrates the nearly bimodal nature of $V_{sw}$ in these observations, i.e., peaks corresponding to slow and fast solar wind streams, with respective averages ~400 km/s and ~620 km/s. For this 70-day period, the separation between the slow and fast solar wind is near ~550 km/s, i.e., where the distribution in Figure 6(a) has a minimum. Given the sampling distribution of the different solar wind speeds, it is not surprising that the 2D-histogram of $dS/dt$ vs. $V_{sw}$ in Figure 6(b) also shows two maxima and a minimum, corresponding to the peaks and the valley seen in Figure 6(a).

We therefore normalize the 2D-histograms to investigate the actual relationship between $dS/dt$ and $V_{sw}$. Figure 6(c) shows the 2D-histogram normalized by the 1D-histogram of $V_{sw}$, which clearly demonstrates that the distribution of entropic rates during this 70-day interval is independent of the solar wind speed. The independence of the entropic rates with respect to $V_{sw}$ is also shown in Figure 6(d). This figure shows the average $dS/dt$ estimated for each of the $V_{sw}$-bins. The 1D-histogram in Figure 6(e) indicates that the mean value from all the bins is ~ −3.22±0.07, which is close to the mode of the slightly asymmetric distribution in Figure 6(d).

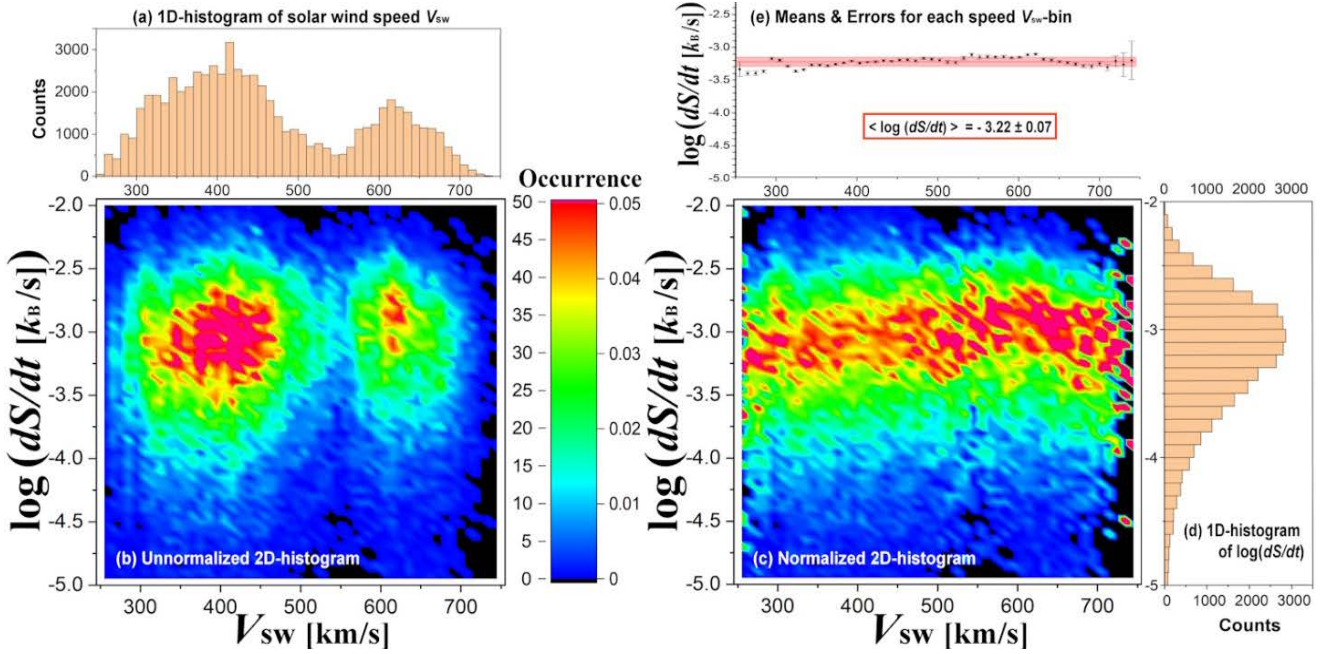

**Figure 6.** (a) 1D-histogram of solar wind speed $V_{sw}$, observed during the 70-day interval and shown in Figure 4a. (b) 2D-histogram of the logarithm of the entropic rates, log($dS/dt$), as shown in Figure 5b, plotted vs. $V_{sw}$. (c) 2D-histogram from panel (b), normalized by the distribution of $V_{sw}$ in panel (a). (d) Mean and standard error of the means for the values of log($dS/dt$) in each $V_{sw}$-bin. (e) Histogram of all values of log($dS/dt$).

The entropic rate is calculated by the difference of sequential entropic values divided by the respective time interval. The entropy is the logarithm of $P/n^{\gamma}$; thus, we approximate numerically the infinitesimal deviation of a logarithmic quantity ln($x$) by the (a) difference of logarithms, $d\ln(x) \approx \ln(x_2)-\ln(x_1) = \ln(x_2/x_1)$, or by the (b) standard difference, $d\ln(x) \approx dx/x \approx (x_2-x_1)/x_1 = x_2/x_1 - 1$. The two approaches are equivalent, as long as $\Delta x$ is small, because of the expansion $\ln(x_2/x_1) = \ln[1+(x_2/x_1-1)] \approx (x_2/x_1-1)$. The time resolution of datasets (~92s) is sufficiently small, so that the entropic rate and turbulent heating, determined by either approximation, lead to similar results; indeed, Figure 7 plots $S_t/n$ in both ways, (a) and (b), while panel (c) gives their percentage difference, $|S_t^{(b)}/S_t^{(a)}-1|$; we observe that the average percentage difference is not larger than ~$10^{-1.5}$~1/30. (Thus, we require all processed $S_t/n$ rates to satisfy $|S_t^{(b)}/S_t^{(a)}-1| < 1/30$.)

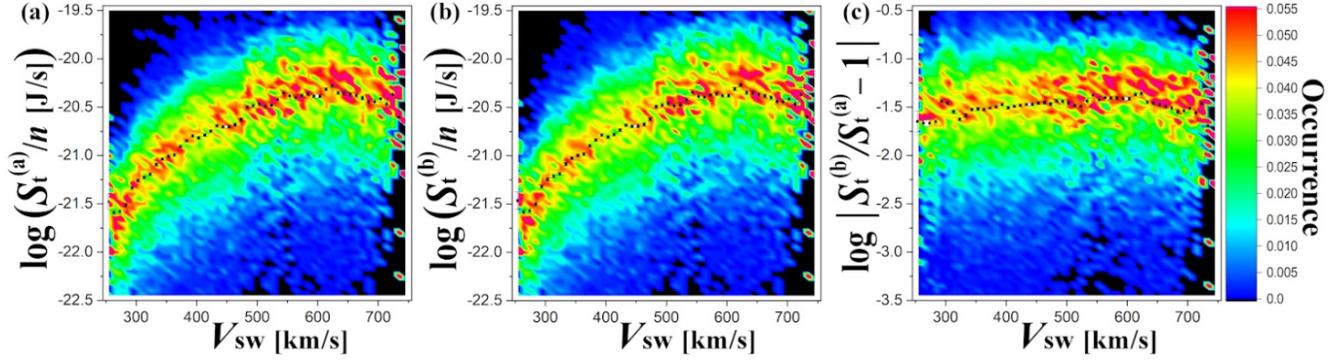

**Figure 7.** 2D-histograms of $S_t/n$ values vs. $V_{sw}$, normalized as explained in Figure 5, and plotted for both approximation ways of (a) "difference of logarithm", (b) "standard difference"; (c) Plot of $|S_t^{(b)}/n - S_t^{(a)}/n|$ normalized by $S_t^{(a)}/n$. All ordinates are plotted by their logarithms.

### 3.2. Relationship between turbulent heating and wave-particle equilibrium energy

The constancy of the ratio of the energy per proton over the plasma frequency, $E_p^0/\omega_{pl}$ (and thus, the concept of plasma particle - wave thermodynamic equilibrium itself), has been previously confirmed by numerous space plasma measurements. The value of the involved Planck-like constant that characterizes stable space plasmas was found to be (Livadiotis 2017; 2019a):

$$\hbar_* = (1.16 \pm 0.08) \times 10^{-22} \, \text{J} \cdot \text{s} \,. \tag{6}$$

The thermodynamic equilibrium characterizing the constancy of $E_p^0/\omega_{pl}$ appears to be violated in the case of the slow solar wind in the inner heliosphere. As has been observed, the ratio $E_p^0/\omega_{pl}$, deviates from the constant value of $\hbar_*$; this deviation is larger for smaller solar wind speeds. Indeed, Figure 8c plots the values of $E_p^0/\omega_{pl}$, using the data shown in Figure 4 and Eq.(1); Figure 9 plots these values of $E_p^0/\omega_{pl}$ vs. solar wind speed $V_{sw}$, where we observe that $E_p^0/\omega_{pl}$ undergoes a continuous transition from the slow to the fast solar wind, tending asymptotically towards the known value of $\hbar_*$. The observed deviation of $E_p^0/\omega_{pl}$ from the constant $\hbar_*$ is caused by a missing energy that was not originally counted in the particle energy; that is, the turbulent energy $E_t$ (Livadiotis 2019). Then, we use the total energy per proton $E_p$ that includes the turbulent energy, $E_p = E_p^0 + E_t$, instead of simply the non-turbulent term $E_p^0$. In this way, the wave-particle equilibrium, $E_p = \hbar_* \cdot \omega_{pl}$, constitutes a fine probe for estimating the turbulent energy: $E_t = \hbar_* \cdot \omega_{pl} - E_p^0$ (as shown in Eq.(1)).

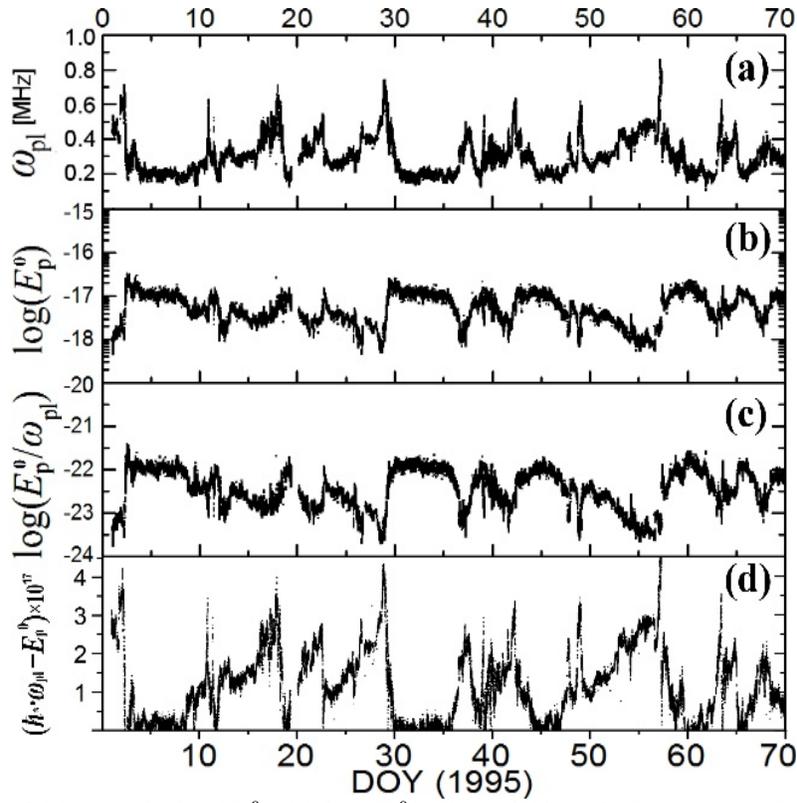

**Figure 8.** Time-series of (a) $\omega_{pl}$, (b) $\log(E_p^0)$, (c) $\log(E_p^0/\omega_{pl})$, and (d) turbulent energy $E_t = \hbar_* \cdot \omega_{pl} - E_p^0$, using the plasma moments datasets during the 70-day period shown in Figure 4.

In Figure 8(c), we observe that there are some time intervals, during which $\log(E_p^0/\omega_{pl})$ reaches a peak value of ~ –22. However, in other periods, $\log(E_p^0/\omega_{pl})$ decreases by more than one order of magnitude, up to ~ –23.5. The existence of strong turbulence determines the conditions where $\log(E_p^0/\omega_{pl})$ deviates from the peak value of $\log \hbar_*$ ~ –22 (Figure 9). Namely, the turbulent energy $E_t$ undergoes a continuous decrease from slow to fast solar wind, tending asymptotically to zero.

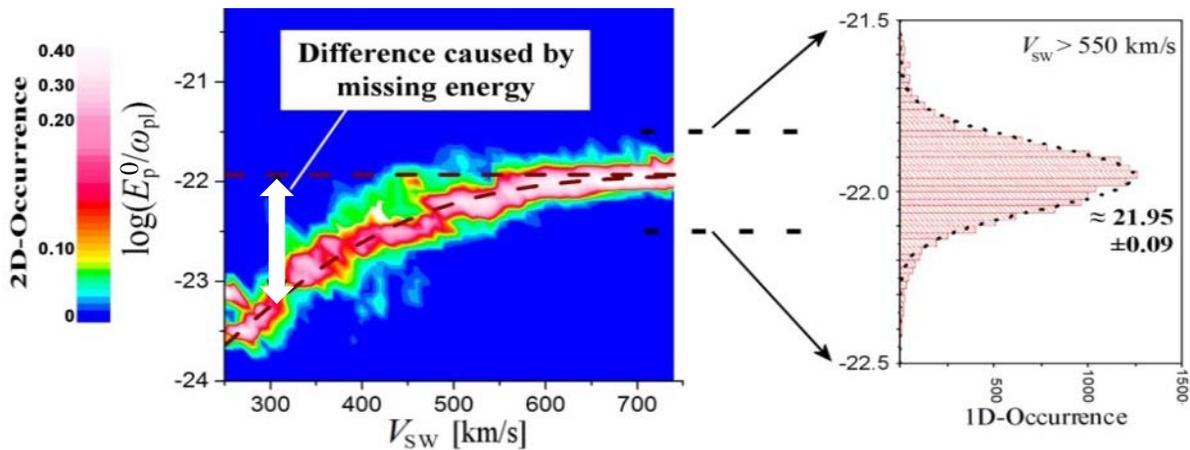

**Figure 9.** Left: 2D-histogram of $\log(E_p^0/\omega_{pl})$ vs. solar wind speed $V_{sw}$, normalized by the 1D-histogram of speeds as explained in Fig.6. Right: 1D-histogram of $\log(E_p^0/\omega_{pl})$ for $V_{sw}>550$ km/s; at such large speeds the ratio $E_p^0/\omega_{pl}$ approaches the value of $\hbar_*$. For lower speeds, $E_p^0/\omega_{pl}$ differs from $\hbar_*$ (horizontal dash line), because of the turbulent

energy (noted as missing energy), which was not originally included in $E_p$. The plots use ~92s solar wind data from Wind S/C during the first 70 day period of 1995, shown in Figure 4. The energy difference, $\hbar_* \cdot \omega_{pl} - E_p^0$, provides the turbulent energy $E_t$ that was not included in the non-turbulent term $E_p^0$ (see Eq.(1)). This missing energy is the turbulent energy heating of the solar wind. (Modified from Livadiotis & Desai 2016)

The radial profiles of this missing energy $\hbar_* \cdot \omega_{pl} - E_p^0$ and the turbulent energy $E_t$ for the solar wind proton plasma in the inner heliosphere were derived and compared in (Figure 10, Livadiotis 2019a). As shown, the energy difference, $\hbar_* \cdot \omega_{pl} - E_p^0$, provides the turbulent energy $E_t$. The connection of the missing plasmon − proton energy with the turbulent energy provides a new method for estimating and cross-examining the turbulent energy in space and astrophysical plasmas.

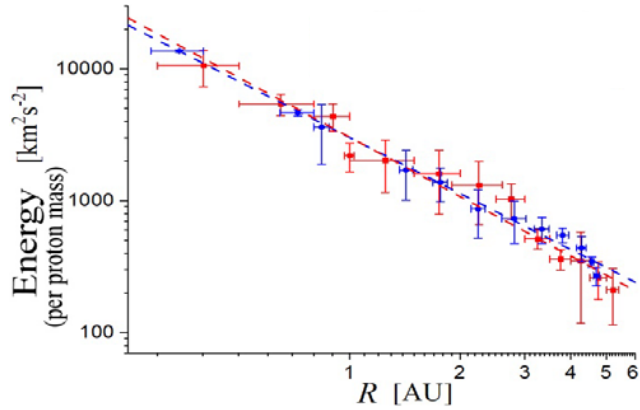

**Figure 10.** Radial profiles in the inner heliosphere of the turbulent energy derived directly from observations (blue) (Adhikari et al. 2015, 2017) and the wave-particle thermodynamic equilibrium (red), characterized by the same statistical model, $E_t(R)/m_p = 10^{3.48 \pm 0.04} \cdot R^{-1.43 \pm 0.07}$. (Taken from Livadiotis 2019a)

Since the paper focuses on $R=R^0 \sim 1$ AU, then, Eq.(3a) becomes $E_t = m_p \cdot E_b^0$, i.e., the 2nd term in Eq.(3a) is negligible. Combining $E_t = m_p \cdot E_b^0 = \hbar_* \cdot \omega_{pl} - E_p^0$ from Eq.(1) and $\lambda_b^0 = m_p \cdot E_b^0 / (S_t/n)$ from Eq.(5), we find the expressions of $E_b^0$ and $\lambda_b^0$ as functions of $\hbar_* \cdot \omega_{pl} - E_p^0$ and $S_t/n$, which can be derived from solar wind plasma moments. Namely,

$$E_b^0 = m_p^{-1}(\hbar_* \cdot \omega_{pl} - E_p^0) \ , \ \lambda_b^0 = m_p^{-\frac{1}{2}}(\hbar_* \cdot \omega_{pl} - E_p^0)^{\frac{3}{2}} \cdot (\tfrac{1}{n} S_t)^{-1} \ . \tag{7}$$

Figure 11 shows the normalized 2D-histrograms of $E_b^0$ and $\lambda_b^0$ determined as shown by the relationships in Eq.(7), and plotted against solar wind speed $V_{sw}$.

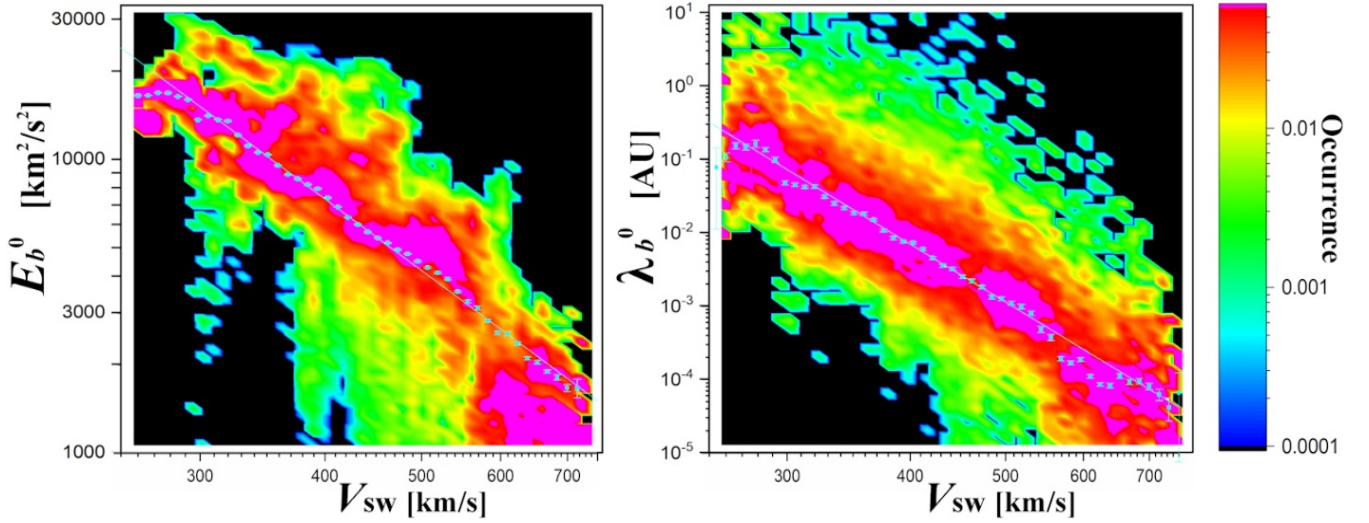

**Figure 11.** 2D-histogram of fluctuating magnetic energy $E_b^0$ (Left) and correlation length $\lambda_b^0$ (Right), plotted against solar wind speed $V_{sw}$, and normalized by the 1D-histogram of speeds as explained in Figure 6(c).

In order to identify the physical conditions during such time periods, we first construct the normalized 2D-histograms by binning the solar wind speed $V_{sw}$ as abscissae and $\log(E_p^0/\omega_{pl})$ as ordinates. As in Figure 6(c), we normalize the raw 2D-histograms by the corresponding 1D-histogram of $V_{sw}$. Then, we generate the dataset of $\hbar_* \cdot \omega_{pl} - E_p^0$, which provides the value of fluctuating magnetic energy $E_b^0$, according to Eq.(7). As shown in Figure 10(a), we construct the normalized 2D-histogram of $E_b^0$ plotted vs. $V_{sw}$, and determine the dependence $E_b^0(V_{sw})$.

This analysis has revealed a brand new feature that was not reported in earlier studies – a remarkably smooth transition from slow to fast wind of the turbulent heating rate, $S_t/n$, as well as, of other turbulence parameters, such as the fluctuating magnetic energy $E_b^0$ and its correlation length $\lambda_b^0$. For instance, $\log(S_t/n)$ transitions smoothly from values of ~ -22 to -20, as the solar wind speed $V_{sw}$ increases from ~300 km/s to ~700 km/s (Figure 7). On the other hand, $E_b^0$ and $\lambda_b^0$ decrease smoothly and linearly with respect to $V_{sw}$ (on a log-log scale), revealing a power-law anti-correlated dependence (all plotted in Figure 10). The best fitted expressions are given in Eq.(8):

$$E_b^0 = A_e \cdot V_{sw}^{-a_e}, \lambda_b^0 = A_\lambda \cdot V_{sw}^{-a_\lambda}: A_e = 10^{10.41 \pm 0.06}, A_\lambda = 10^{19.02 \pm 0.22}; a_e = 2.515 \pm 0.025, a_\lambda = 8.14 \pm 0.08, \quad (8)$$

with units: $E_b^0$ in [km/s]$^2$, $\lambda_b^0$ in [AU], and $V_{sw}$ in [km/s].

### 3.3. Variations of turbulent heating parameters with solar wind speed

We can use the results of §3.1 and §3.2 to understand the effects of the embedded solar wind and ICMEs (or other interplanetary structures) on the derived turbulence parameters, $S_t/n$, $E_b^0$, $\lambda_b^0$.

As an example, we demonstrate the variations of the turbulence parameters between the solar wind and an ICME (observed for ~24h at doy-29 of the 70-day period shown in Figure 4). Figure 12 shows the turbulence parameters of the ICME compared to those of solar wind. Clearly, the correlation length, describing the size of the biggest eddies in the turbulent plasma, remains invariant; also, the ICME's turbulent heating is enhanced, with consequences in space plasma processes and thermodynamics (Verma et al. 1995; Vasquez et al. 2007; Sorriso-Valvo et al. 2007; Marino et al. 2008; Elliott et al. 2019; Livadiotis 2019b). For instance, the temperature rate $dT/dt$ and the exponent $\xi$ of the radial temperature profile ($T \sim R^{-\xi}$) can be connected with the turbulent heating rate $S_t/n$, so that $dT/dt$ is larger and/or $\xi$ slightly smaller in the examined ICME, rather than in the solar wind.

A similar future analysis of the full ensemble of the structures over two solar cycles will provide insights into the physical processes interwoven with the variations of the turbulence parameters.

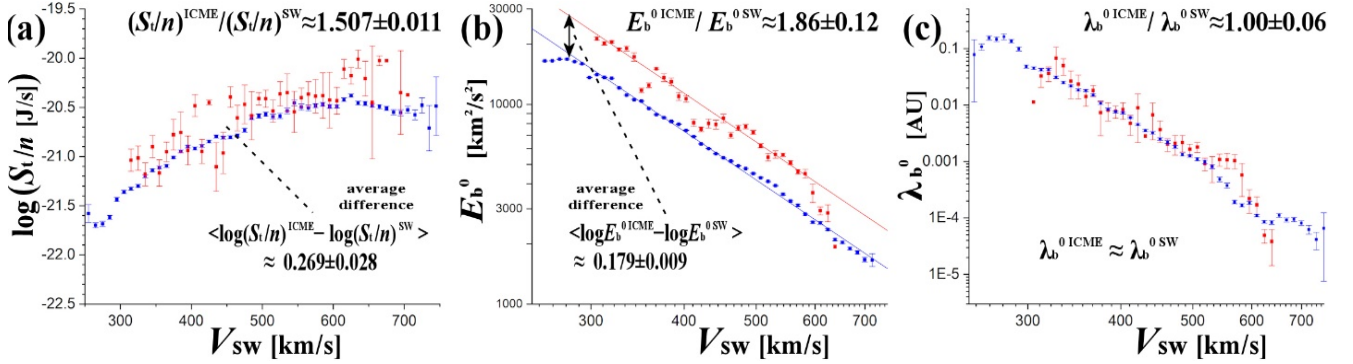

**Figure 12.** Plots of turbulence parameters (a) turbulent heating rate, $S_t/n$, (b) fluctuating magnetic energy $E_b^0$, and (c) its respective correlation length $\lambda_b^0$, for both the solar wind and an ICME plasma. The variations of the parameters between solar wind and ICME plasma are shown.

## 4. Conclusions

We employed two physical properties of the expanding solar wind plasma, (a) wave-particle thermodynamic equilibrium, and (b) transport of entropic rate, for improving our understanding of the turbulent heating of solar wind at 1AU.

We analyzed plasma moments and field datasets from Wind S/C, in order to compute (i) the fluctuating magnetic energy $E_b^0$, (ii) the corresponding correlation length $\lambda_b^0$, and (iii) the turbulent heating rate $S_t/n$. We then identified their relationships against the solar wind speed, as well as, the variation of these relationships when compared for the plasmas of solar wind and an ICME.

The wave-particle thermodynamic equilibrium requires the energy of a plasmon, that is, the quantum of plasma oscillation, to be balanced by the energy per proton, which sums the magnetic energy, enthalpy, and turbulent energy; we then express the energy equation with the only unknown, the turbulent energy, which provides a good approximation of the fluctuating magnetic energy $E_b^0$.

On the other hand, the transport of entropy and its rate involves finding the turbulent heating rate, $S_t/n$, which is expressed in terms of $E_b^0$ and the corresponding correlation length $\lambda_b^0$. Our methodology uses the wave-particle thermodynamic equilibrium to compute $E_b^0$, and it is combined with the transport of entropy to compute $S_t/n$ and $\lambda_b^0$.

In summary, the paper results are outlined as follows:
1) A new method is developed for deriving turbulent heating using plasma moments and magnetic field data;
2) We computed the turbulent heating and its related parameters, i.e., (i) the fluctuating magnetic energy $E_b^0$, (ii) the corresponding correlation length $\lambda_b^0$, and (iii) the turbulent heating rate $S_t/n$;
3) We construct the solar wind speed dependence of the turbulent heating and its related parameters;

4) The turbulent heating rate, $S_t/n$, was shown to increase with increasing solar wind speed for the slow wind, and tending towards a constant for the fast wind.

5) An empirical power-law relationship of decreasing $E_b^0$ and $\lambda_b^0$ with increasing solar wind speed was found, where the involved exponents were rather sharp, i.e., $\sim -2.5$ and $\sim -8$, respectively.

6) An ICME's turbulent heating rate and fluctuating magnetic energy are enhanced when compared with the solar wind plasma, while the correlation length remains about invariant.

The analysis showed how the turbulent heating and its related parameters can be computed using simply the first plasma moments and the magnetic field; this work made it straightforward that a similar analysis to the full ensemble of the interplanetary structures over two solar cycles will provide insights into the physical processes interwoven with the turbulent heating rate and related parameters.

Authors MAD and GL acknowledge partial support from NASA grants 80NSSC19K0079 and 80NSSC20K0290.

**Appendix: Uncertainty toolbox**

The uncertainties are derived from propagating the errors of the involved variables and parameters; e.g., the uncertainty of $Y$ that depends on $\{X_i\}$, $i=1,2,\ldots$ is $\delta Y = [\sum_i (\partial Y/\partial X_i)^2]^{1/2}$.

The 2nd row of Table 1 shows the propagated (logarithmic) uncertainties of: plasma frequency $\omega_{pl}$, thermal pressure $P$, proton plasma beta $\beta$, entropy $S$ ($k_B$ units), entropic rate $\Delta S/\Delta t$), turbulent heating rate $S_t/n$, non-turbulent $E_p^0$ and fluctuating magnetic $E_b^0$ energies, correlation length $\lambda_b^0$.

The plasma density, temperature, and IP magnetic field strength have log-normal distributed errors, which, on average, are $\delta \ln n = \delta n/n \approx 3\%$, $\delta \ln T = \delta T/T \approx 8\%$ (Kasper et al. 2006), and $\delta \ln B = \delta B/B \sim 1\%$ (Lepping et al. 1995; Farrell et al. 1996) also $\langle \Delta t \rangle \sim 92s$. These are used for computing the approximations of 3rd row of Table 1.

**Table 1. Uncertainties of derived quantities**

| $X$ | $\delta X$ | $\langle \delta X \rangle \approx$ |
|---|---|---|
| $\log \omega_{\text{pl}\,i}$ | $\frac{1}{2}(\delta \log n_i)$ | 0.0065 |
| $\log P_i$ | $\sqrt{(\delta \log n_i)^2 + (\delta \log T_i)^2}$ | 0.037 |
| $\log \beta_i$ | $\sqrt{(\delta \log n_i)^2 + (\delta \log T_i)^2 + 4(\delta \log B_i)^2}$ | 0.038 |
| $S_i / k_{\text{B}}$ | $\ln(10)\sqrt{(\delta \log n_i)^2 + 2.25(\delta \log T_i)^2}$ | 0.12 |
| $\Delta(S_i / k_{\text{B}}) / \Delta t_i$ | $[\ln(10)/\Delta t_i]\sqrt{(\delta \log n_i)^2 + (\delta \log n_{i+1})^2 + 2.25[(\delta \log T_i)^2 + (\delta \log T_{i+1})^2]}$ | $0.175 / \Delta t_i \cong 0.0019$ |
| $\log(\frac{1}{n_i} S_{\text{t}i})$ | $(1-q_i)^{-1}\sqrt{(\delta \log n_i)^2 + (\delta \log n_{i+1})^2 + (0.5 + q_i)^2 (\delta \log T_i)^2 + 2.25(\delta \log T_{i+1})^2}$ | $0.08(1-q_i)^{-1}\sqrt{q_i^2 + q_i + 2.78}$ |
| $\log E_{\text{p}\,i}^0$ | $(1+2.5\beta_i)^{-1}\sqrt{(\delta \log n_i)^2 + 4(\delta \log B_i)^2 + 6.25\beta_i^2(\delta \log T_i)^2}$ | $0.2(1+2.5\beta_i)^{-1}\sqrt{0.03 + \beta_i^2}$ |
| $\log E_{\text{b}\,i}^0$ | $(\epsilon_i - \beta_i^{-1} - 2.5)^{-1}\sqrt{(\frac{1}{2}\epsilon_i - \beta_i^{-1})(\delta \log n_i)^2 + 4\beta_i^{-2}(\delta \log B_i)^2 + 6.25(\delta \log T_i)^2 + \epsilon_i^2(\delta \log \hbar_*)^2}$ | $0.0087(\epsilon_i - \beta_i^{-1} - 2.5)^{-1}$ $\times\sqrt{100 + 2.25(\frac{1}{2}\epsilon_i - \beta_i^{-1}) + \beta_i^{-2} + 5.3\epsilon_i^2}$ $\cong 0.09(\epsilon_i - \beta_i^{-1} - 2.5)^{-1}$ |
| $\log \lambda_{\text{b}\,i}^0$ | $\sqrt{2.25(\delta \log E_{\text{b}\,i}^0)^2 + [\delta \log(\frac{1}{n_i} S_{\text{t}i})]^2}$ | $0.08\sqrt{(1-q_i)^{-2}(q_i^2 + q_i + 2.78) + 2.85(\epsilon_i - \beta_i^{-1} - 2.5)^{-2}}$ |
| Definitions | $q_i \equiv s_i / s_{i+1}$, $s_i \equiv T_i^{\frac{3}{2}} n_i^{-1}$, $\Delta t_i \equiv t_{i+1} - t_i$, $\beta_i = 2\mu_0 n_i k_{\text{B}} T_i / B_i^2$, $\epsilon_i \equiv \hbar \omega_{\text{pl}\,i} / (k_{\text{B}} T_i)$ | |

**ORCID iDs**

G. Livadiotis https://orcid.org/0000-0002-7655-6019
M. A. Dayeh https://orcid.org/0000-0001-9323-1200
G. P. Zank https://orcid.org/0000-0002-4642-6192